\documentclass[prd,aps]{revtex4}

\begin{document}
\title{Modern space-time and undecidability}

\author{Rodolfo Gambini}
\address{Instituto de F\'{\i}sica, Facultad de Ciencias, 
Universidad
de la Rep\'ublica, Igu\'a 4225, CP 11400 Montevideo, Uruguay}
\author{Jorge Pullin}
\address{Department of Physics and Astronomy, 
Louisiana State University, Baton Rouge,
LA 70803-4001}

\date{December 5th. 2007}

\begin{abstract}
        The picture of space-time that Minkowski created in 1907 has
        been followed by two important developments in physics not
        contained in the original picture: general relativity and
        quantum mechanics. We will argue that the use of concepts of
        those theories to construct space-time implies conceptual
        modifications in quantum mechanics. In particular one can
        construct a viable picture of quantum mechanics without a
        reduction process that has outcomes equivalent to a picture
        with a reduction process. One therefore has two theories that
        are entirely equivalent experimentally but profoundly
        different in the description of reality they give. This
        introduces a fundamental level of undecidability in physics of
        a kind that has not been present before. We discuss some of
        the implications.
\end{abstract}

\maketitle
\section{Introduction}

In 1907 Minkowski noted that the natural arena to formulate the
special theory of relativity of Einstein was space-time. In his
address delivered at the 80th Assembly of German Natural Scientists
and Physicians in 1908, Minkowski remarks: {\it ``The views of space
and time which I wish to lay before you have sprung from the soil of
experimental physics, and therein lies their strength. They are
radical. Henceforth space by itself, and time by itself, are doomed to
fade away into mere shadows, and only a kind of union of the two will
preserve an independent reality.''}. ``The soil of experimental
physics'' in 1907 differs from the one today. At that time it was
thought that space and time were continuous and that they could be
measured with arbitrary precision. This was altered when quantum
mechanics was developed in the 1930's. There it was noted that
fundamental uncertainties were germane to modern physics. In 1957
Salecker and Wigner \cite{wigner} decided to revise the picture that
Minkowski had painted of space-time introducing concepts of quantum
mechanics. In particular they considered limitations in the accuracy
of clocks and rulers that one needs in order to construct the picture
of space-time that we are familiar with. But fundamental uncertainties
in quantum mechanics can be minimized. The situation changes
dramatically when general relativity is brought into the picture in
two different ways. On the one hand general relativity leads to
fundamental uncertainties that cannot be minimized. In particular one
cannot measure distances and times beyond a minimum level of
uncertainty, as was emphasized by Ng and Van Dam \cite{ng}. The second
way in which general relativity alters the picture is that space-time
is now a dynamical arena that is not directly observable. The only
observable properties of space-time are relational in nature.
Therefore unlike the space-time of Minkowski, which was an immutable
arena for special relativity, in general relativity space-time becomes
an object that cannot be directly probed and the properties of it that
can be probed are relational in nature and subject to fundamental
minimum uncertainties.

To further elaborate the point, consider the measurement of time. 
In its usual formulation, quantum mechanics involves an idealization.
The idealization is the use of a perfect classical clock to measure
times. Such a device clearly does not exist in nature, since all
measuring devices are subject to some level of quantum fluctuations.
Therefore the equations of quantum mechanics, when cast in terms of
the variable that is really measured by a clock in the laboratory,
will differ from the traditional Schr\"odinger description. Although
this is an idea that arises naturally in ordinary quantum mechanics,
it is of paramount importance when one is discussing quantum gravity.
This is due to the fact that general relativity is a generally
covariant theory where one needs to describe the evolution in a
relational way. One ends up describing how certain objects change when
other objects, taken as clocks, change. At the quantum level this
relational description will compare the outcomes of measurements of
quantum objects. Quantum gravity is expected to be of importance in
regimes (e.g. near the big bang or a black hole singularity) in which
the assumption of the presence of a classical clock is clearly
unrealistic. The question therefore arises: is the difference between
the idealized version of quantum mechanics and the real one just of
interest in situations when quantum gravity is predominant, or does it
have implications in other settings? 

We will argue that indeed it does have wider implications.  Some of
them are relevant to conceptual questions (e.g.  the black hole
information paradox or ultimate limitations on quantum computing) and
there might even be experimental implications.  A detailed discussion
of these ideas can be found in previous papers
\cite{cqg,njp,piombino}, and in particular in the pedagogical review
\cite{obregon}. Here we present an abbreviated discussion as an
introduction to a remarkable consequence of these ideas: that the
nature of physical processes in modern physics becomes {\em
undecidable}.  In a nutshell we observe that when one considers
fundamental limitations in the measurements of space-time, unitary
quantum mechanics behaves in the same way as quantum mechanics with a
reduction process.  The two theories therefore become equivalent in
their physical predictions, yet they are radically different in
nature: in one of them quantum states are determined once and for all
at the beginning of the universe and are not changed by physical
events.  In the other the picture is dramatically different, with
quantum states constantly changing as events produces measurable
outcomes. In spite of these differences, there is no experimental way
to decide which of these pictures corresponds to reality. The
implications of this observation philosophically are profound.  They
advocate what is technically known in philosophical circles as a
``regularist'' point of view towards the nature of physical laws,
drastically different from the ``necessitarian'' point of view that
was considered at the time of Minkowski's conception of space-time as
the most natural interpretation of the role of physical laws.

The plan of this paper is as follows: in the next section we will
discuss the form of the evolution equation of quantum mechanics when
the time variable, used to describe it, is measured by a real clock.
In section III we will consider a fundamental bound on how accurate
can a real clock be and the implications it has for quantum mechanics
in terms of real clocks and its consequences. Section IV discusses 
briefly some possible experimental implications of the proposal. 
Section V discusses conceptual implications in the foundations
of quantum mechanics. Section VI argues that the conceptual
implications leads to a new notion of undecidability in the nature
of physical processes in quantum theory. We end with a discussion.

\section{Quantum mechanics with real clocks}

Given a physical situation we start by choosing a ``clock''. By this
we mean a physical quantity (more precisely a set of quantities, like
when one chooses a clock and a calendar to monitor periods of more
than a day) that we will use to keep track of the passage of {\it
  time}. An example of such a variable could be the angular position
of the hand of an analog watch. Let us denote it by $T$. We then
identify some physical variables that we wish to study as a function
of time.  We shall call them generically $O$ (``observables'').
We then proceed to quantize the system by promoting all the
observables and the clock variable to self-adjoint quantum operators
acting on a Hilbert space. The latter is defined once a well defined
inner product is chosen in the set of all physically allowed states.
Usually it consists of squared integrable functions $\psi(q)$ with
$q$ the configuration variables.

Notice that we are not in any way modifying quantum mechanics. We
assume that the system has an evolution in terms of an external
parameter $t$, which is a classical variable, given by a Hamiltonian and
with operators evolving with Heisenberg's equations (it is easier to
present things in the Heisenberg picture, though it is not mandatory
to use it for our construction). Then the standard rules of quantum 
mechanics and its probabilistic nature apply.

We will call the eigenvalues of the ``clock'' operator $T$ and the
eigenvalues of the ``observables'' $O$.  We will assume that the clock
and the measured system do not interact (if one considered an
interaction it would produce additional effects to the one discussed).
So the density matrix of the total system is a direct product of that
of the system under study and the clock $\rho =\rho_{\rm sys} \otimes
\rho_{\rm cl}$, and the system evolves through a unitary evolution
operator that is of the tensor product form $U=U_{\rm sys}\otimes
U_{\rm cl}$. The quantum states are described by a density matrices at
a time $t$. Since the latter is unobservable, we would like to shift
to a description where we have density matrices as functions of the
observable time $T$.  We define the probability that the resulting
measurement of the clock variable $T$ correspond to the value $t$,
\begin{equation}
{\cal P}_t(T) \equiv {{\rm Tr}\left(P_T(0) U_{\rm cl}(t)\rho_{\rm cl} U_{\rm cl}(t)^\dagger\right)\over
\int_{-\infty}^\infty dt\,{\rm Tr}\left(P_T(t) \rho_{\rm cl}\right)},
\end{equation}
where $P_T(0)$ is the projector on the eigenspace with eigenvalue $T$
evaluated at $t=0$. We 
note that $\int_{-\infty}^\infty dt {\cal P}_t(T)=1$. We now
define the evolution of the density matrix,
\begin{equation}
\rho(T) \equiv \int_{-\infty}^\infty dt U_{\rm sys}(t) \rho_{\rm sys} 
 U_{\rm sys}(t)^\dagger {\cal P}_t(T)
\end{equation}
where we dropped the ``sys'' subscript in the left hand side since
it is obvious we are ultimately interested in the density matrix of
the system under study, not that of the clock.

We have therefore ended with an ``effective'' density matrix in the
Schr\"odinger picture given by $\rho(T)$. It is possible to reconstruct
entirely in a relational picture the probabilities using this effective
density matrix, for details we refer the reader to the lengthier discussion
in \cite{obregon}. By its very definition, it
is immediate to see that in the resulting evolution unitarity is lost,
since one ends up with a density matrix that is a superposition of
density matrices associated with different $t$'s and that each evolve
unitarily according to ordinary quantum mechanics.

Now that we have identified what will play the role of a density
matrix in terms of a ``real clock'' evolution, we would like to see
what happens if we assume the ``real clock'' is behaving
semi-classically. To do this we assume that ${\cal P}_t(T) =f(T-T_{\rm
  max}(t))$, where $f$ is a function that decays very rapidly for
values of $T$ far from the maximum of the probability distribution
$T_{\rm max}$. We refer the reader to \cite{obregon} for a derivation,
but the resulting evolution equation for the probabilities is (in the limit
in which corrections are small),
\begin{equation}
{\partial \rho(T)\over \partial T} =i [\rho(T),H] +\sigma(T) [H,[H,\rho(T)]].
\end{equation}
and the extra term is dominated by the rate of change $\sigma(T)$ of
the width of the distribution $f(t-T_{\rm max}(t))$.

An equation of a form more general than this has been considered in
the context of decoherence due to environmental effects, it is called
the Lindblad equation. Our particular  form of the equation is such
that conserved quantities are automatically preserved by the modified
evolution.  Other mechanisms of decoherence coming from a different
set of effects of quantum gravity have been criticized in the past
because they fail to conserve energy \cite{hag}. It should be noted
that Milburn arrived at a similar equation as ours from different
assumptions \cite{milburn}. Egusquiza, Garay and Raya derived a
similar expression from considering imperfections in the clock due to
thermal fluctuations \cite{egusquiza}. It is to be noted that such
effects will occur in addition to the ones we discuss here.
Corrections to the Schr\"odinger equation from quantum gravity have
also been considered in the context of WKB analyses
\cite{kiefersingh}. Considering time as a quantum
variable has also been discussed by Bonifacio \cite{Bo} with
a formulation somewhat different than ours but with similar
conclusions.

Another point to be emphasized, particularly in the context of a volume
celebrating Minkowski's contributions, is that our approach has been quite
naive in the sense that we have kept the discussion entirely in terms
of non-relativistic quantum mechanics with a unique time across space.
It is clear that in addition to the decoherence effect we discuss
here, there will also be decoherence spatially due to the fact that
one cannot have clocks perfectly synchronized across space and also
that there will be fundamental uncertainties in the determination of
spatial positions.  This, together with the issue of the Lorentz invariance of the predictions,  is discussed in some detail in our paper
\cite{spatial}.

\section{Fundamental limits to realistic clocks}

We have established that when we study quantum mechanics with a
physical clock (a clock that includes quantum fluctuations), unitarity
is lost, conserved quantities are still preserved, and pure states
evolve into mixed states. All this in spite of the fact that the underlying
theory is unitary as usual. The effects are more pronounced the worse
the clock is.  Which raises the question: is there a fundamental
limitation to how good a clock can be? This question was first
addressed by Salecker and Wigner \cite{wigner}. Their reasoning went
as follows: suppose we want to build the best clock we can. We start
by insulating it from any interactions with the environment.  An
elementary clock can be built by considering a photon bouncing between
two mirrors. The clock ``ticks'' every time the photon strikes one of
the mirrors. Such a clock, even completely isolated from any
environmental effects, develops errors. The reason for them is that by
the time the photon travels between the mirrors, the wavefunctions of
the mirrors spread. Therefore the time of arrival of the photon
develops an uncertainty.  Salecker and Wigner calculated the
uncertainty to be $\delta t \sim \sqrt{t/M}$ where $M$ is the mass of
the mirrors and $t$ is the time to be measured (we are using units
where $\hbar=c=1$ and therefore mass is measured in 1/second). The
longer the time measured the larger the error. The larger the
mass of the clock, the smaller the error.

So this tells us that one can build an arbitrarily accurate clock just
by increasing its mass. However, Ng and Van Dam \cite{ng} pointed
out that there is a limit to this. Basically, if one piles up enough
mass in a concentrated region of space one ends up with a black hole.
Some readers may ponder why do we need to consider a concentrated
region of space. The reason is that if we allow the clock to be more
massive by making it bigger, it also deteriorates its performance
(see the discussion in
\cite{ngotro} in response to \cite{baez}).
  
A black hole can be thought of as a clock since it is an oscillator.
In fact it is the ``fastest'' oscillator one can have, and therefore
the best clock for a given size.   It has normal modes of
vibration that have frequencies that are of the order of the light
travel time across the Schwarzschild radius of the black hole.  (It is
amusing to note that for a solar sized black hole the frequency is in
the kilohertz range, roughly similar to that of an ordinary bell). The
more mass in the black hole, the lower the frequency, and therefore
the worse its performance as a clock. This therefore creates a tension
with the argument of Salecker and Wigner, which required more mass to
increase the accuracy. This indicates that there actually is a ``sweet
spot'' in terms of the mass that minimizes the error. Given a time to
be measured, light traveling at that speed determines a distance, and
therefore a maximum mass one could fit into a volume determined by
that distance before one forms a black hole.  That is the optimal
mass.  Taking this into account one finds that the best accuracy one
can get in a clock is given by $\delta T \sim T_{\rm Planck}^{2/3}
T^{1/3}$ where $t_{\rm Planck}=10^{-44}s$ is Planck's time and $T$ is
the time interval to be measured. This is an interesting result. On
the one hand it is small enough for ordinary times that it will not
interfere with most known physics. On the other hand is barely big
enough that one might contemplate experimentally testing it, perhaps
in future years.

With this absolute limit on the accuracy of a clock we can quickly
work out an expression for the $\sigma(T)$ that we discussed in the
previous section \cite{prl,piombino}. It turns out to be $\sigma(T)=
\left({T_{\rm Planck}}\over{T_{\rm max}-T}\right)^{1/3}T_{\rm
  Planck}$. With this estimate of the absolute best accuracy of a
clock, we can work out again the evolution of the density matrix for a
physical system in the energy eigenbasis. One gets
n\begin{equation}
  \rho(T)_{nm} = \rho_{nm}(0) 
e^{-i\omega_{nm} T} e^{-\omega_{nm}^2  T_{\rm Planck}^{4/3} T^{2/3}}.
\end{equation}

So we conclude that {\em any} physical system that we study in the lab
will suffer loss of quantum coherence at least at the rate given by
the formula above. This is a fundamental inescapable limit. A pure
state inevitably will become a mixed state due to the impossibility of
having a perfect classical clock in nature.

\section{Possible experimental implications}

Given the conclusions of the previous section, one can ask what are
the prospects for detecting the fundamental decoherence we propose.
At first one would expect them to be dim. It is, like all quantum
gravitational effects, an ``order Planck'' effect. But it should be
noted that the factor accompanying the Planck time can be rather
large. For instance, if one would like to observe the effect in the
lab one would require that the decoherence manifest itself in times of
the order of magnitude of hours, perhaps days at best. That requires
energy differences of the order of $10^{10}eV$ in the Bohr frequencies
of the system. Such energy differences can only be achieved in
``Schr\"odinger cat'' type experiments, but are not outrageously
beyond our present capabilities.  Among the best candidates today are
Bose--Einstein condensates, which can have $10^6$ atoms in coherent
states. However, it is clear that the technology is still not there to
actually detect these effects, although it could be possible in
forthcoming years.

A point that could be raised is that atomic clocks currently have an
accuracy that is less than a decade of orders of magnitude worse than
the absolute limit we derived in the previous section. Couldn't
improvements in atomic clock technology actually get better than our
supposed absolute limit? This seems unlikely. When one studies in
detail the most recent proposals to improve atomic clocks, they
require the use of entangled states \cite{atomic} that have to remain
coherent. Our effect would actually prevent the improvement of atomic
clocks beyond the absolute limit!

Finally, if one has doubts about the effect's existence, one
must recall that one can make it arbitrarily large just by picking
a lousy clock. This is of course, not terribly interesting and is
not what is normally done in physics labs. 
But it should be noted that experiments of Rabi oscillations in
rubidium atoms measure certain correlations which can be interpreted
as having the atom work as a lousy clock. The oscillations show
experimentally the exponential decay we discuss. See Bonifacio
et al. \cite{bonifacioetal} for a discussion. 

\section{Conceptual implications}

The fact that pure states evolve naturally into mixed states has 
conceptual implications in at least three interesting areas of physics.
The first two we have discussed before so we only mention them for
completeness and to refer the reader to previous papers:
the black hole information paradox \cite{prl,piombino} and
quantum computation \cite{qc}. The third area where the effects we
discussed could be of interest are in foundational issues in 
quantum mechanics, in particular, the measurement problem.
We will now expand a bit on what we mean by this.

The measurement problem in quantum mechanics is
related to the fact that in ordinary quantum mechanics the measurement
apparatus is assumed to be always in an eigenstate after a measurement
has been performed.  The usual explanation \cite{Schlossauser} for
this is that there exists interaction with the environment. This
selects a preferred basis, i.e., a particular set of quasi-classical
states that commute, at least approximately, with the Hamiltonian
governing the system-environment interaction. Since the form of the
interaction Hamiltonians usually depends on familiar ``classical''
quantities, the preferred states will typically also correspond to the
small set of ``classical'' properties. Decoherence then quickly damps
superpositions between the localized preferred states when only the
system is considered. This is taken as an explanation of the
appearance to a local observer of a ``classical'' world of
determinate, ``objective'' (robust) properties.

There are two  main problems with such a point of view. The first one
is how is one to
interpret the local suppression of interference in spite of the
fact that the total state describing the system-environment
combination retains full coherence. One may raise the question 
whether retention of the full coherence could ever lead to 
empirical conflicts with the ascription of definite values to 
macroscopic systems. The usual point of view is that it would
be very difficult to reconstruct the off diagonal elements of
the density matrix in practical circumstances. However, at least
as a matter of principle, one could indeed reconstruct such
terms (the evolution of the whole system remains unitary
\cite{omnes}) by ``waiting long enough''. The second problem
is that even if the state ends up being ``quasi-diagonal'' 
in the preferred basis of the measurement device,  this is not
necessarily a completely satisfactory solution to the measurement
problem. This is known as the ``and-or'' problem. As Bell \cite{Bell}
put it ``if one were not actually on the lookout for probabilities,
... the obvious interpretation of even $\rho'$ [the reduced density
matrix] would be that the system is in a state in which various
$|\Psi_m>$'s coexist:
\begin{equation}
|\Psi_1><\Psi_1|\quad {\rm and}\quad |\Psi_2><\Psi_2|
\quad {\rm and} \quad \ldots
\end{equation}
This is not at all a {\it probability} interpretation, in which the
different terms are seen not as {\it coexisting} but as {\it
alternatives}.''  We will return on this point when we introduce
our idea of undecidability later in this paper.

Our mechanism of fundamental decoherence could contribute to the
understanding of the two problems mentioned above. In the usual
system-environment interaction the off-diagonal terms of the density
matrix oscillate as a function of time. Since the environment is
usually considered to contain a very large number of degrees of
freedom, the common period of oscillation for the off-diagonal terms
to recover non-vanishing values is very large, in many cases larger
than the life of the universe. This allows to consider the problem
solved in practical terms. When one adds in the effect we discussed,
since it suppresses exponentially the off-diagonal terms, one never
has the possibility that the off-diagonal terms will see their initial
values restored, no matter how long one waits.

To analyze the implications of the use of real clocks in the
measurement problem, we will discuss in some detail an example. In
spite of the universality of the loss of coherence we introduced, it
must be studied in specific examples of increasing level of
realism. The simplest example we can think of is due to Zurek
\cite{zurek}.  This simplified model does not have all the effects of
a realistic one, yet it exhibits how the information is transferred
from the measuring apparatus to the environment. The model consists of
taking a spin one-half system that encodes the information about the
microscopic system plus the measuring device. A basis in its two
dimensional Hilbert space will be denoted by $\{|+>,|->\}$. The
environment is modeled by a bath of many similar two-state systems
called atoms. There are $N$ of them, each denoted by an index $k$ and
with associated two dimensional Hilbert space $\{|+>_k,|->_k\}$. The
dynamics is very simple, when there is no coupling with the
environment the two spin states have the same energy, which is taken
to be $0$.  All the atoms have zero energy as well in the absence of
coupling. The whole dynamics is contained in the coupling, given by
the following interaction Hamiltonian
\begin{equation}
H_{\rm int} = \hbar \sum_k \left( g_k \sigma_z \otimes \sigma_z^k \otimes
\prod_{j\neq k} I_j\right).
\end{equation}
In this notation $\sigma_z$ is analogous to a Pauli spin matrix.  It
has eigenvalues $+1$ for the spin eigenvector $|+>$ and $-1$ for
$|->$; it acts as the identity operator on all the atoms of the
environment. The operators $\sigma^k_z$ are similar, each acts like a
Pauli matrix on the states of the specific atom $k$ and as the
identity upon all the other atoms and the spin. $I_j$ denotes the
identity matrix acting on atom $j$ and $g_k$ is the coupling constant
that has dimensions of energy and characterizes the coupling energy of
one of the spins $k$ with the system. In spite of the abstract
character of the model, it can be thought of as providing a sketchy
model of a photon propagating in a polarization analyzer.

Starting from a normalized initial state
\begin{equation}
|\Psi(0)> = \left(a|+> + b|->\right) \prod_{k=1}^N \otimes \left[
\alpha_k|+>_k +\beta_k |->_k \right],
\end{equation}
it is easy to solve the Schroedinger equation and one gets for the
state at the time $t$,
\begin{eqnarray}
|\Psi(t)> \label{psit}
&=& a |+> \prod_{k=1}^N\otimes\left[
\alpha_k\exp\left(ig_k t\right)|+>_k 
+ \beta_k \exp\left(-ig_k t\right)|->\right]\\
&&+ b |-> 
\prod_{k=1}^N\otimes\left[
\alpha_k\exp\left(-ig_k t\right)|+>_k 
+ \beta_k \exp\left(ig_k t\right)|->\right].\nonumber
\end{eqnarray}

Writing the complete density operator $\rho(t) = |\Psi(t)><\Psi(t)|$,
one can take its trace over the environment degrees of freedom to get
the reduced density operator,
\begin{equation}
\rho_c(t) = |a|^2 |+><+| + |b|^2 |-><-| + z(t) ab^* |+><-| 
+z^*(t) a^* b|-><+|,
\end{equation}
where
\begin{equation}
z(t) = \prod_{k=1}^N \left[\cos\left(2g_k t\right)+i\left(|\alpha_k|^2
-|\beta_k|^2\right) \sin\left(2 g_k t\right)\right].
\end{equation}

The complex number $z(t)$ controls the value of the non-diagonal
elements. If this quantity vanishes the reduced density matrix
$\rho_c$ would correspond to a totally mixed state (``proper
mixture''). That would be the desired result, one would have several
classical outcomes with their assigned probabilities. However,
although the expression we obtained vanishes quickly assuming the
$\alpha$'s and $\beta$'s take random values, it behaves like a
multiperiodic function, i.e. it is a superposition of a large number
of periodic functions with different frequencies. Therefore this
function will retake values arbitrarily close to the initial value for
sufficiently large times. This implies that the apparent loss of
information about the non-diagonal terms reappears if one waits a long
enough time. This problem is usually called ``recurrence of
coherence''.  The characteristic time for these phenomena is
proportional to the factorial of the number of involved
frequencies. Although this time is usually large, perhaps exceeding
the age of the universe, at least in principle it implies that the
measurement process does not correspond to a change from a pure to a
mixed state in a fundamental way.

The above derivation was done using ordinary quantum mechanics in
which one assumes an ideal clock is used to measure time. If one
re-does the derivation using the effective equation we derived for
quantum mechanics with real clocks one gets the same expression for
$z(t)$ except that it is multiplied by $\prod_k\exp\left(-(2 g_k)^2
T^{4/3}_{\rm Planck} t^{2/3}\right)$.  That means that asymptotically
the off diagonal terms indeed vanish, the function $z(t)$ is not
periodic anymore. Although the exponential term decreases slowly with
time, the fact that there is a product of them makes the effect quite
relevant. 

Therefore we see that the inclusion of real clocks makes impossible
the observation of ``revivals'' in systems where the measurement
process leads to observable outcomes. The observation of 
``revivals'' not only is very difficult to observe in a practical sense
due to the length of time that elapses between revivals, but becomes
unobservable due to fundamental reasons, irrespective of the 
advanced level of the technology used for the observations. As a consequence
the revivals are not a means to determine if global coherence
in the total system has been preserved or if there has been a change
in the state of the measurement system.

Revivals are not the only possible manifestation of the total
unitary evolution of the system when an observation of a given
result is made in the system. Other experiments have been
proposed. These are also of considerable practical difficulty since
they involve measurements of the total system. For example, 
d'Espagnat \cite{despagnat} has proposed a technique based on
the observation of constant of the motion of the total system.
Such constants of the motion take different values if the reduction
process has happened or not. These experiments require the
construction of macroscopic ensembles including the environment
prepared in the same state and subsequent measurements of 
each of the degrees of freedom involved. The concrete models
proposed \cite{despagnat} for these types of experiments are
highly idealized, ignoring interactions between the degrees of 
freedom of the environment and their individual evolution. The
free Hamiltonians of the environment and the measuring apparatus
is assumed zero. In these models the loss of unitarity due to
real clocks appears to be insufficient to eliminate all possibilities
of distinguishing between a reduction and a unitary evolution in
an observable process due to the fact that one would be dealing
with a constant of the motion. Preliminary analyses of concrete 
implementations of these ideas in more realistic models lead us
to believe that the loss of spatial coherence and dispersion 
during evolution of the wavefunctions of the component degrees
of freedom and the back reaction of the measuring device on
the system \cite{spekkens} plus the evolution of the measuring device appear to
imply that the observation of the relevant observables is 
impossible. We are currently studying some of these models
and will report elsewhere on the details.

\section{Undecidability and the laws of physics}

We have shown that quantum mechanics with and without an explicit
mechanism for state reduction yield the same physical predictions.
There are good reasons to consider that it is experimentally
impossible to distinguished between two alternative situations, the
one resulting from a unitary evolution of the system plus the
environment and the other one where the system plus the measuring
device undergo an abrupt change of state.  There are therefore two
complete logical structures that yield the same predictions and
therefore one cannot decide experimentally which one ``corresponds to
reality''. This is what we refer to as ``undecidability''.

The concept of undecidability in physics was also discussed by 
Wolfram \cite{wolfram} in a different context, although not totally
disconnected from the one we introduce here. He refers to the
undecidability of computational nature about the outcome of a 
physical process. There exist physical processes whose outcome
is not predictable. The optimum computational process to determine
the outcome is the physical process itself. In this sense we could
speak of {\em undecidability in the outcome} of a process. This 
type of undecidability is weaker than the one we are discussing,
since the issues can be decided just by waiting till the outcome
occurs. The undecidability we are referring to does not refer to 
the outcome of a process but to the nature of the process.

In philosophy there are different attitudes that have been taken towards
the physical laws of nature (see for instance \cite{stanford}). 
One of them is the ``regularity theory''; in it, the laws of physics are
statements about uniformities or regularities of the world and therefore
are just ``convenient descriptions'' of the world. The laws of physics
are dictated by a preexisting world and are a representation of our
ability to describe the world. Another point of view is the ``Necessitarian
theory''. There laws of nature are ``principles'' which govern the natural
phenomena, that is, the world ``obeys'' the laws of nature. The presence
of the undecidability we point out suggests strongly that the ``regularity
theory'' point of view is more satisfactory.

If one takes seriously the regularity point of view one can ponder
about the nature of reality. Does the physical world have a reduction
process, does it not, or does it depend on the case?  In the case in
which there is no reduction process, in the Heisenberg picture the
state of a system is given and eternal. If there is a reduction
process the state changes every time there is an event resulting in a
measurement. The third possibility, which is suggested by the
undecidability, is that the system may choose between behaving as if
there is a reduction process or not. This would add to the well known
probabilistic freedom of quantum systems characterized by the free
will theorem of Conway and Kochen \cite{conwaykochen}
 another one characterized by two
alternative behaviors in relation to the rest of the universe. That
is, after the observation of the event either the system simply
behaves as if it were part of the universe and its state were that of
the universe or if as its state would be given by the reduction
postulate. In the first case the system would keep its entanglement
with the rest of the universe (i.e. the environment), in the second it
will lose its entanglement.

If one adopts what is probably the most attractive view that considers
that the universe always evolves unitarily and therefore quantum
states are determined once and for all no matter what is the chosen
behavior of the subsystem under observation one needs to face the
problem of when do events happen in such a framework. Our point of
view is that an event occurs when the experimental distinction between
coexisting or exclusionary alternatives becomes undecidable, since in
that instant the predictions of the laws of physics are not altered by
the possible reduction of the state of the system associated with the
information acquired when the event takes place.



\section{Discussion}

We have argued that the use of realistic clocks in quantum mechanics
implies that pure states evolve into mixed states in spite of the fact that
the underlying equations of quantum mechanics are unitary. The effect is
further compounded if one allows realistic rulers to measure distances
and the formulation can be made Lorentz covariant. This alters
fundamentally the picture of space-time as a static fixed entity that
was first introduced by Minkowski. Conceptually, it also has profound
implications since it allows to construct a quantum mechanics without
the introduction of a reduction process that nevertheless has the same
predictions as ordinary quantum mechanics with a reduction
process. The nature of physical processes is therefore undeterminate
in a novel and fundamental way that adds itself to other previous
proposals of undecidability in physics.

\section{Acknowledgements}

This work was supported in part by grants NSF-PHY0650715,
and by funds of the Horace
C. Hearne Jr. Institute for Theoretical Physics, FQXi, PEDECIBA (Uruguay)
and CCT-LSU.


\begin{references}

\bibitem{wigner}
E. Wigner, Rev. Mod. Phys. 29, 255 (1957).
\bibitem{ng}
Y.~J.~Ng and H.~van Dam,
Annals N.\ Y.\ Acad.\ Sci.\  {\bf 755}, 579 (1995) 
[arXiv:hep-th/9406110];
Mod.\ Phys.\ Lett.\ A {\bf 9}, 335 (1994); see also
F. K\'arolhy\'azy, A. Frenkel, B. Luk\'acs in ``Quantum
concepts in space and time'' R. Penrose and C. Isham, editors,
Oxford University Press, Oxford (1986).
\bibitem{cqg}
R. Gambini, R. Porto, J. Pullin,
Class.\ Quant.\ Grav.\  {\bf 21}, L51 (2004)
[arXiv:gr-qc/0305098].

\bibitem{njp} R. Gambini, R. Porto, J. Pullin,
New J.\ Phys.\  {\bf 6}, 45 (2004)
[arXiv:gr-qc/0402118].

\bibitem{piombino} R. Gambini, R. Porto, J. Pullin, 
  Braz.\ J.\ Phys.\  {\bf 35}, 266 (2005)
  [arXiv:gr-qc/0501027].

\bibitem{obregon}
  R.~Gambini, R.~Porto and J.~Pullin,
  Gen.\ Rel.\ Grav.\  {\bf 39}, 1143 (2007)
  [arXiv:gr-qc/0603090].

\bibitem{hag} See for instance J. Ellis, J. Hagelin, D.V. Nanopoulos,
and M. Srednicki, Nucl. Phys. B241 (1984) 381; T. Banks, M.E. Peskin,
and L. Susskind, Nucl. Phys. B244 (1984) 125.

\bibitem{milburn} G. J. Milburn, Phys. Rev {\bf A44}, 5401 (1991).

\bibitem{egusquiza}
I. Egusquiza, L. Garay, J. Raya, Phys. Rev. {\bf
A59}, 3236 (1999) 
[arXiv:quant-ph/9811009].

\bibitem{kiefersingh} C. Kiefer, T. Singh, Phys. Rev. {\bf D44}, 1061 (1991).

\bibitem{Bo} R. Bonifacio, Nuo. Cim. {\bf
D114}, 473 (1999).  

\bibitem{spatial}
R.~Gambini, R.~A.~Porto and J.~Pullin,
  Int.\ J.\ Mod.\ Phys.\  D {\bf 15}, 2181 (2006)
  [arXiv:gr-qc/0611148].


\bibitem{ngotro} 
  Y.~J.~Ng and H.~van Dam,
  Class.\ Quant.\ Grav.\  {\bf 20}, 393 (2003)
  [arXiv:gr-qc/0209021].

\bibitem{baez}
J.~C.~Baez and S.~J.~Olson,
  Class.\ Quant.\ Grav.\  {\bf 19}, L121 (2002)
  [arXiv:gr-qc/0201030].

\bibitem{prl} 
R.~Gambini, R.~A.~Porto and J.~Pullin,
  Phys.\ Rev.\ Lett.\  {\bf 93}, 240401 (2004)
  [arXiv:hep-th/0406260].


\bibitem{atomic} See for instance A. Andre, A. Sorensen, M. Lukin,
  Phys. Rev. Lett {\bf 92}, 230801 (2004) [arXiv:quant-ph/0401130].


\bibitem{bonifacioetal} R. Bonifacio {\em et al.} Phys.
Rev. {\bf A61}, 053802 (2000).

\bibitem{qc} R. Gambini, R. Porto, J. Pullin, in
``Gravity, astrophysics and strings at the black sea'', P. Fiziev,
M. Todorov, editors, St. Kliment Ohridski Press (2006)
[arXiv:quant-ph/0507262].  



\bibitem{Schlossauser} M. Schlosshauer, Rev. Mod. Phys. 
{\bf 76}, 1267 (2004) [arXiv:quant-ph/0312059].

\bibitem{omnes} R. Omn\`es, ``The interpretation of quantum mechanics'', 
Princeton Series in Physics, Princeton, NJ (1994).

\bibitem{Bell} J. Bell ``Against 'Measurement''' in ``Sixty two years of
uncertainty'', A. Miller (editor), Plenum, New York (1990).


\bibitem{zurek} W. Zurek, Phys. Rev. {\bf D26},
1862 (1982).  



\bibitem{despagnat} B. d'Espagnat, ``Veiled reality'',
Westview, Boulder, Colorado (2003).  

\bibitem{spekkens} S. Bartlett, T. Rudolph, R. Spekkens, P. Turner,
New J. Phys. {\bf 8}, 58 (2006); D. Poulin, J. Yard, New J. Phys.
{\bf 9} 156 (2007).

\bibitem{wolfram}
S. Wolfram, Phys. Rev. Lett. {\bf 54},  735 (1985).

\bibitem{stanford} See for instance ``Laws of Nature'' in 
The Stanford encyclopedia of Philosophy (http://plato.stanford.edu)
and references therein.


\bibitem{conwaykochen} J. Conway, S. Kochen, Found. Phys. {\bf 36}, 
1441 (2006).

\end{references}
\end{document}